# SOLAR WIND ACCELERATION IN CORONAL HOLES

**Steven R. Cranmer**

Harvard-Smithsonian Center for Astrophysics, Cambridge, MA 02138, USA
*Email:* scranmer@cfa.harvard.edu

ABSTRACT

This paper reviews the current state of our understanding of high-speed solar wind acceleration in coronal holes. Observations by *SOHO,* coupled with interplanetary particle measurements going back several decades, have put strong constraints on possible explanations for how the protons, electrons, and minor ions receive their extreme kinetic properties. The asymptotic plasma conditions of the wind depend on energy and momentum deposition *both* at the coronal base (where, e.g., the mass flux is determined) and in the extended acceleration region between 2 and 10 solar radii (where the plasma becomes collisionless and individual particle species begin to exhibit non-Maxwellian velocity distributions with different moments). The dissipation of magnetohydrodynamic fluctuations (i.e., waves, turbulence, and shocks) is believed to dominate the heating in the extended corona, and spectroscopic observations from the UVCS instrument on *SOHO* have helped to narrow the field of possibilities for the precise modes, generation mechanisms, and damping channels. We will survey recent theoretical and observational results that have contributed to new insights, and we will also show how next-generation instruments can be designed to identify and characterize the dominant physical processes to an unprecedented degree.

Key words: coronal holes; solar corona; solar wind; plasma physics; turbulence; UV spectroscopy.

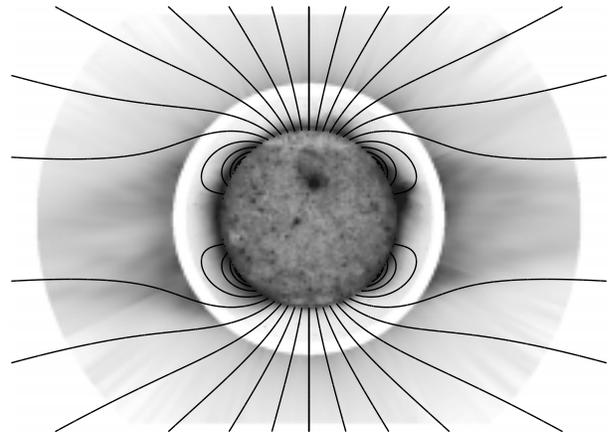

*Figure 1. The solar corona on 17 August 1996 (near solar minimum), with bright regions plotted as dark. The inner image is the solar disk in Fe XII 195 Å emission from EIT/SOHO. The outer image is the extended corona in O VI 1032 Å emission from UVCS/SOHO. The axisymmetric field lines are from the solar-minimum model of Banaszkiewicz et al. (1998).*

## 1. INTRODUCTION

Coronal holes were first observed by Waldmeier (1957), at the Swiss Federal Observatory in Zürich, as long-lived regions of negligible intensity in coronagraphic images of the 5303 Å green line (see also Waldmeier, 1975). Coronal holes were effectively re-discovered in the early 1970s as discrete dark patches on the X-ray and ultraviolet solar disk. In the late 1960s and early 1970s, it was realized that coronal holes extend into interplanetary space as large-scale open magnetic regions that coincide with the high-speed component of the solar wind (e.g., Wilcox, 1968; Krieger et al., 1973; Nolte et al., 1976; Zirker, 1977). At the minimum of the Sun's 11-year activity cycle, large coronal holes exist at the north and south heliographic poles and extend into a large fraction of the volume of the heliosphere (see Figure 1). At times other than solar minimum, smaller and more transient coronal holes appear at all latitudes, with plasma properties intermediate between those of polar coronal holes and the higher-density regions of the corona (see Wang & Sheeley 1990; Miralles et al., 2001a, 2001b; and Miralles, 2002, these proceedings).

This paper provides a brief review of the physics of coronal holes and the acceleration of the associated high-speed solar wind. The *Solar and Heliospheric Observatory* (*SOHO*) spacecraft has led to a dramatic increase in our understanding of coronal holes and the solar wind, and this paper focuses mainly on insights gleaned from *SOHO* observations. There are many broader reviews of these topics that should be consulted for a more complete picture; see, for example, Parker (1963, 1991, 2001), Hundhausen (1972), Leer et al. (1982, 1998), Pneuman (1986), Barnes (1992), Tu & Marsch (1995), Esser & Habbal (1997), Feldman & Marsch (1997), Marsch (1999), Cranmer (2001a, 2002a, 2002b), Velli (2001), and Hollweg & Isenberg (2002).

## 2. OBSERVATIONS

Any hypotheses concerning the origin of coronal heating and solar wind acceleration must be tested by comparing the predicted plasma properties with observations. The two most useful means of measuring these properties have been *in situ* spacecraft detection and the remote sensing of coronal photons. Some key results of these measurements—listed in order of increasing distance from the Sun—are summarized below.

Instruments aboard *Yohkoh, TRACE* (*Transition Region And Coronal Explorer*), and *SOHO* have revealed strong variability and complexity at the base of the corona on the smallest observable scales. The SUMER instrument on *SOHO* has investigated the origins of the high-speed solar wind in coronal holes by mapping out blueshifts in coronal emission lines (Hassler et al., 1999; Peter & Judge, 1999). SUMER measurements have also shown that ion temperatures exceed electron temperatures at very low heights (Seely et al., 1997; Tu et al., 1998). Obtaining reliable electron temperatures above the limb ($\sim$1.1 to 1.4 $R_\odot$), though, has proved difficult, with inferred values ranging between $3 \times 10^5$ and $1.7 \times 10^6$ K at comparable heights (see Ko et al., 1997; David et al., 1998; Aschwanden & Acton, 2001; Doschek et al., 2001). The reconciliation of this controversy may be the existence of non-Maxwellian electron distributions at low coronal heights (Esser & Edgar, 2000), but there may also be selection effects due to different instrumental sensitivities in an intrinsically multi-thermal distribution of temperatures.

Remote-sensing measurements of plasma properties in the acceleration region of the solar wind (i.e., 2 to 10 $R_\odot$) require the bright solar disk to be occulted. The invention of the coronagraph by Lyot in the 1930s led to the ability to observe the extended corona at all times. Subsequently, the development of spaceborne ultraviolet coronagraph spectrometers in the 1970s (Kohl et al., 1978; Withbroe et al., 1982) led to the ability to perform detailed studies of the kinetic properties of atoms, ions, and electrons. The UVCS instrument aboard *SOHO* provided the first measurements of ion temperature anisotropies and differential outflow speeds in the acceleration region of the wind (Kohl et al., 1997, 1998, 1999). UVCS measured $O^{5+}$ perpendicular temperatures exceeding $10^8$ K at heights above 2 $R_\odot$ (see Figure 2), with the anisotropy ratio $T_\perp / T_\parallel$ reaching values of order 10 to 100. Temperatures for both $O^{5+}$ and $Mg^{9+}$ are significantly greater than mass-proportional when compared to hydrogen, and outflow speeds for $O^{5+}$ may exceed those of hydrogen by as much as a factor of two (see also Li et al., 1998; Cranmer et al., 1999b; Giordano et al., 2000).

Spacecraft have measured *in situ* particle velocity distribution functions and electromagnetic fields as close to the Sun as 60 $R_\odot$ (e.g., *Helios 1* and *2*), and as far out as 12,000 $R_\odot$ (e.g., *Voyager 2*). Velocity distributions contain information about the macroscopic plasma properties (density, flow velocity, temperature, heat flux) as well as the physics on smaller kinetic scales. For example, the anisotropic cores of proton distributions in the high-speed wind (see Marsch et al., 1982) signal the preferential de-position of energy perpendicular to the background magnetic field. Continued heating in interplanetary space is also inferred by the radial dependence of the two ("core" and "halo") components of the electron velocity distribution (e.g., Phillips et al., 1995). Most heavy ions flow faster than the bulk proton-electron wind by about 5% to 10% (e.g., Hefti et al., 1998), and their temperatures exceed simple mass-proportionality with protons in the highest-speed flows (Collier et al., 1996). It is noteworthy that the UVCS measurements discussed above are similar in character to the *in situ* data, but they imply more extreme departures from thermodynamic equilibrium in the corona that grow progressively weaker with increasing heliocentric distance.

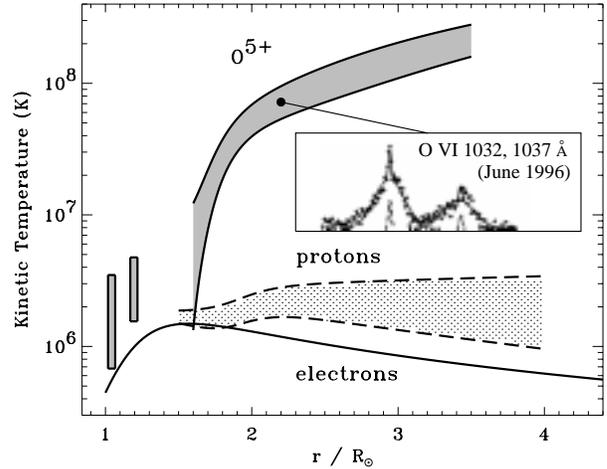

*Figure 2. Summary plot of coronal hole kinetic temperature measurements. Perpendicular temperatures for protons and $O^{5+}$ above 1.5 $R_\odot$ are from an empirical model that reproduced UVCS line widths (Kohl et al., 1998; Cranmer et al., 1999b). The two $O^{5+}$ boxes at lower heights are representative of ion temperatures derived from SUMER line widths (e.g., Hassler et al., 1997), and the electron temperature is from Ko et al. (1997). Additional uncertainties, mainly due to differences between plumes and interplume regions, are not shown here.*

## 3. PROPOSED PHYSICAL PROCESSES

The energy that heats the corona and accelerates the solar wind must originate in subphotospheric convective motions, but the physical processes that transport this energy to the corona and convert it into thermal, kinetic, and magnetic energy are not known. Different physical mechanisms for heating the corona probably govern loops, bright points, and the large-scale open field lines of coronal holes (e.g., Narain & Ulmschneider, 1990, 1996; Priest et al., 2000). There is also a growing realization that the coronal base ($r < 1.5 R_\odot$) is probably heated by different processes than those that apply at larger heliocentric distances. This heuristic division is supported by the drastic differences in Coulomb collision rates at the base (where all species seem to be collisionally coupled) and in the supersonic wind (which is nearly collisionless). The two regimes are also differentiated by the complex-ity and topology of the magnetic field (e.g., Dowdy et al., 1986), and by the strength of Spitzer-Härm heat conduction (which channels the basal heat downwards into a sharp transition region, but is extremely weak in the extended corona; see Li, 1999; Lie-Svendsen et al., 2002).

The remainder of this paper discusses extended heating in the wind acceleration region. It is now generally believed that the high-speed solar wind cannot be produced without the existence of gradual energy deposition above the base of the corona (see, e.g., Leer et al., 1982; Parker 1991; Barnes et al., 1995; Hollweg, 1999; see however, Lemaire & Pierrard, 2001). The vast majority of theoretical models of the extended corona involve the transfer of energy from propagating magnetic fluctuations (i.e., waves, shocks, or turbulence) to the particles. This general consensus arises because the ultimate source of energy must be solar in origin, and thus it must somehow *propagate* out to the distances where the heat deposition must occur.

It is not known how or where the fluctuations responsible for extended coronal heating are generated. Alfvén waves have received the most attention because they seem to be the least damped by collisional processes (i.e., viscosity, conductivity, resistivity), and thus are most likely to escape the chromosphere and transition region. However, there have been recent observations that imply the presence of slow magnetosonic waves in coronal holes at large heights (e.g., Ofman et al., 1999). At distances greater than 2–3 $R_\odot$, wave dissipation should begin to be dominated by collisionless processes. From an empirical perspective, the most likely damping mechanism seems to be ion cyclotron resonance, since Landau damping tends to preferentially heat electrons in a low-$\beta$ plasma (Habbal & Leer, 1982).

High-frequency (10 to $10^4$ Hz) ion cyclotron waves have been suggested to be generated both at the coronal base (Axford & McKenzie 1992; McKenzie et al., 1995; Tu & Marsch, 1997; Ruzmaikin & Berger, 1998) and in the extended corona by, e.g., turbulent cascade or plasma instabilities (Isenberg & Hollweg, 1983; Hollweg, 1986; Hu et al., 1999; Matthaeus et al., 1999; Markovskii, 2001). Neither paradigm is free from troubling questions. Base-generation of ion cyclotron waves may occur, but Cranmer (2000, 2001b) showed that there exist a significant number of minor ions that can absorb a basal fluctuation spectrum before any primary plasma constituents (protons, $He^{2+}$, or even $O^{5+}$) can be resonantly heated in the extended corona (see also Tu & Marsch, 2001). In addition, Hollweg (2000) predicted that a base-generated cyclotron resonant wave spectrum would be inconsistent with observed radio-scintillation density fluctuations.

Gradual generation of ion cyclotron waves in the extended corona seems to be more likely than base generation, but there are problems with this scenario as well. Ion cyclotron frequencies in the corona are typically 10 to $10^4$ Hz, but the oscillation frequencies observed on the surface of the Sun (generated mainly by convection) are of order $10^{-2}$ Hz. Any means of transferring power from the dominant population of MHD waves to ion cyclotron waves must therefore bridge a gap of many orders of magnitude in frequency (see Figure 3). Most mod-els of MHD turbulence favor the transfer of energy from small to large perpendicular wavenumbers (i.e., fluctuations propagating transversely to the background magnetic field; see, e.g., Shebalin et al., 1983; Goldreich & Sridhar, 1997). Alfvénic fluctuations having large $k_\perp$ and small $k_\parallel$ do not have high frequencies approaching the cyclotron resonances. Observations of *in situ* solar wind fluctuations support the view that high-$k_\perp$ modes dominate (Matthaeus et al., 1990; Bieber et al., 1996), but there does seem to be evidence for high-$k_\parallel$ modes that dissipate via ion gyroresonance (e.g., Leamon et al., 1998). Current work is focusing on the rich variety of generation and dissipation mechanisms that arise for waves propagating obliquely to the background field.

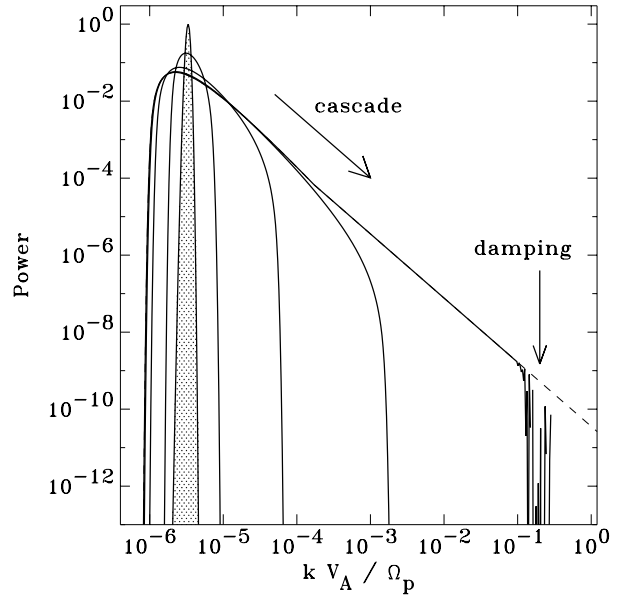

*Figure 3. Schematic cartoon depicting the turbulent cascade of wave power from large MHD scales (i.e., small wavenumbers) to small kinetic scales (i.e., large wavenumbers), where damping occurs. The intermediate steps in the cascade were computed using the diffusion formalism of Zhou & Matthaeus (1990), and the damping rates from Cranmer (2000) were applied at large wavenumbers. The existence of a power-law turbulent spectrum in the $k_\parallel$ direction is an open question.*

## 4. FUTURE DIAGNOSTICS

*SOHO* has provided significant insight into the physics of the acceleration region of the solar wind, but the basic physical processes (e.g., the precise wave generation and dissipation channels) have not yet been identified with certainty. New observations are required to make further progress. For example, next-generation space-based coronagraph spectrometers are being designed with the capability to sample the velocity distributions of dozens of ions in the acceleration region of the high-speed wind. In addition to being sensitive to many more emission lines, such instruments could also detect subtle departures from Gaussian line shapes that signal the presence of specific non-Maxwellian distributions (e.g., Cranmer, 1998, 2001b). Higher spatial and temporal resolution would allow a much more detailed census of mass, momentum, and energy to be performed in the highly filamentary corona. For example, coronal holes have been typically interpreted as a two-phase medium—i.e., consisting of dense polar plumes surrounded by a lower density interplume plasma. There may be, however, a continuous spectrum of density variations in coronal hole flux tubes rather than just two separate phases. New diagnostics such as the measurement of the Thomson scattered H I Ly$\alpha$ profile (which probes the line-of-sight electron velocity distribution; see Fineschi et al., 1998) can put firm constraints on both the "core" electron temperature and the existence of halo-like or power-law wings with unprecedented confidence limits.

In order to illustrate how measuring a larger number of emission line widths would better constrain models of extended coronal heating, let us utilize a simple analytic description of ion cyclotron resonance. Ignoring Coulomb collisions and adiabatic cooling, the perpendicular most-probable speed $w_{\perp i}$ of ion species $i$ evolves according to

$$\frac{dw_{\perp i}^2}{dt} = \frac{\pi \Omega_i^2}{B_0^2} V_{\rm ph} P_B(k_{\rm res}) \qquad (1)$$

where $\Omega_i$ is the ion Larmor frequency, $B_0$ is the background magnetic field strength, $P_B$ is the magnetic fluctuation spectrum of parallel-propagating Alfvén waves, and $k_{\rm res}$ is the parallel wavenumber at ion cyclotron resonance (e.g., Hollweg 1999; Cranmer 2000). For cold plasma dispersion, the Alfvén wave phase speed (in the frame of the wind) at resonance is given by

$$V_{\rm ph} = \frac{\Omega_i}{k_{\rm res}} = V_A (1 - \zeta)^{1/2} \qquad (2)$$

where $\zeta$ is the dimensionless charge-to-mass ratio (i.e., $Z_i/A_i$ in units of the proton charge and mass) and $V_A$ is the Alfvén speed.

For a power-law dependence of the form $P_B \propto k_{\rm res}^{-\eta}$, the heating rate on the right-hand side of equation (1) is proportional to $\zeta^{2-\eta}(1 - \zeta)^{(1+\eta)/2}$. This provides a simple prediction for the $\zeta$ dependence of ion cyclotron heating, but UVCS observations imply that the picture is more complicated. At $r \approx 2\,R_\odot$ in polar coronal holes, the line widths of O$^{5+}$ and Mg$^{9+}$ were of order 400 and 220 km/s, respectively (Kohl et al., 1999). Since these two ions have values of $\zeta$ that are relatively similar (i.e., 0.31 and 0.37), the large observed difference in their widths was not expected. Two possible explanations have been proposed. Esser et al. (1999) suggested that stronger collisional coupling with the protons would lead to lower Mg$^{9+}$ temperatures. Cranmer (2000) noticed that Mg$^{9+}$ has almost the same value of $\zeta$ as O$^{6+}$, the third most abundant ion in the corona. Thus, if ion cyclotron waves are damped in proportion to their relative number densities, there may be substantially lower wave power at the Mg$^{9+}$ resonance than at the O$^{5+}$ resonance.

Observing a larger number of ions can resolve the ambiguity between the above two possible explanations. Let us model the effects of Coulomb collisions and resonant wave damping, and predict the line widths for a number of ions that a next-generation coronagraph spectrometer should be able to observe.

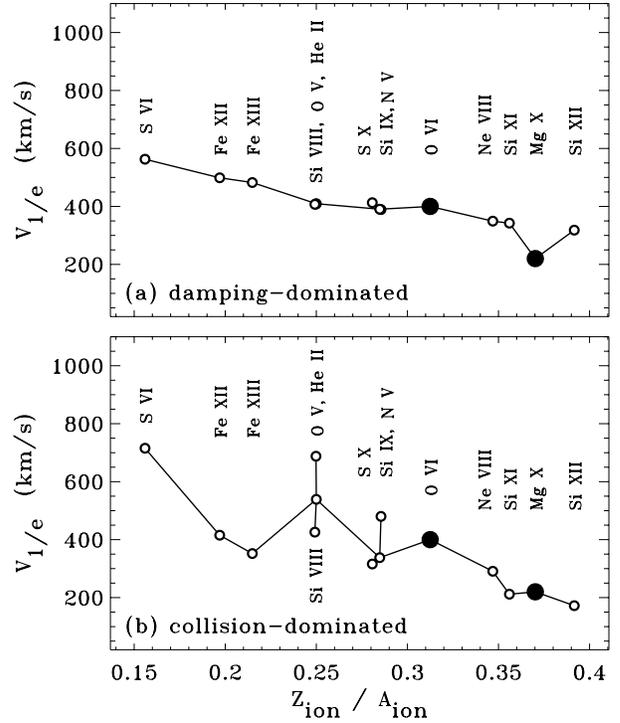

*Figure 4. Estimated line widths for a number of ion species observable with next-generation coronal spectroscopy. The top panel assumes that cyclotron damping is the dominant cause of the narrower Mg X lines (compared to O VI) observed by UVCS, and the bottom panel assumes that Coulomb collisions are responsible.*

Coulomb collisional energy exchange between a given minor ion species and protons could be a significant "temperature drain" on the ions. Let us first compute the ion perpendicular heating from equation (1), then apply the simple temperature equilibration formula

$$\frac{dT_{\perp i}}{dt} = \nu(T_p - T_{\perp i}) \,, \qquad (3)$$

which for a known collision rate $\nu$ has the analytic solution

$$T_{\perp i}(t) = T_p + [T_{\perp i}(0) - T_p]\,e^{-\nu t} \,. \qquad (4)$$

The solar wind expansion time $t_{\rm exp}$ (over which the wind flows through a density scale height) is an appropriate order-of-magnitude value to use for $t$. For O$^{5+}$ in a coronal hole, the product $\nu t_{\rm exp}$ decreases rapidly, from a value of $\sim$2.5 at 1.5 $R_\odot$, to $\sim$0.1 at 2 $R_\odot$, to $\sim$0.02 at 2.5 $R_\odot$ (see Figure 6a of Cranmer et al., 1999a). The collision rate $\nu$ depends on ion charge and mass via the quantity $Z_i^2/A_i$ (Spitzer, 1962).

Ion heating via ion cyclotron resonance also damps the waves in the $P_B$ spectrum. When the waves in the spectrum propagate up from low heights, they enter into cyclotron resonance when their wind-frame frequencies become of the same order as $\Omega_i$. Cranmer (2000) showed that this occurs in a finite damping zone that extracts a fraction of the incoming power $P_{B,0}$,

$$P_B(k_{\rm res}) \approx P_{B,0} \, \exp\left(-\frac{2\gamma_i L_\Omega \, w_{\|i}}{V_{\rm ph}^2}\right) \quad (5)$$

where $\gamma_i$ is the maximum (resonant) value of the linear damping rate (i.e., imaginary part of the frequency), $L_\Omega = \Omega_i/|\partial\Omega_i/\partial r|$ is a representative scale height for the radial variation of $\Omega_i$, and $w_{\|i}$ is the ion parallel most-probable speed. Cranmer (2000, 2001b) concluded, however, that the wave power in the extended corona should be replenished in part by some as-yet-unknown mechanism; this implies that the effective $\gamma_i$ should be reduced by some factor; in the model described below, we multiply the computed value of $\gamma_i$ by a constant factor (less than 1) to take heuristic account of whatever wave generation mechanism is competing with the resonant damping. (This of course is no substitute for a self-consistent calculation of, e.g., simultaneous turbulent cascade and damping as depicted in Figure 3.)

Figure 4 shows the results of computing ion line widths using a steady-state, finite-difference representation of equation (1), i.e.,

$$\frac{dw_{\perp i}^2}{dt} \approx u_{\|i} \frac{\Delta(w_{\perp i}^2)}{\Delta r} \quad (6)$$

with $\Delta r = 1\,R_\odot$ centered around a height of $2\,R_\odot$, and $u_{\|i} \approx 200$ km/s. The adopted power-law spectral exponent $\eta$ is 2.5. The observed $1/e$ line width $V_{1/e}$ is determined according to

$$V_{1/e} \;=\; \sqrt{\xi^2 + \left[w_{\rm o}^2 + \frac{\Delta r}{u_{\|i}}\left(\frac{dw_{\perp i}^2}{dt}\right)\right]} \quad (7)$$

where $\xi = 75$ km/s is a representative low-frequency wave broadening factor (see, e.g., Esser et al., 1999) and $w_{\rm o}$ is the modeled ion most-probable speed at the lower boundary of the finite-difference zone (we assume the ions are cool at $1.5\,R_\odot$, with $w_{\rm o} \approx 30$ km/s).

Figure 4a shows the result of ignoring Coulomb collisions and adjusting both the wave power normalization and the constant multiplier of $\gamma_i$ until both $O^{5+}$ and $Mg^{9+}$ achieve their observed widths. Figure 4b shows the result of ignoring wave damping (i.e., assuming $\gamma_i = 0$) and adjusting the normalization of the Coulomb collision rate $\nu$ until $O^{5+}$ and $Mg^{9+}$ achieve their observed widths. In actuality, of course, both damping and collisions could be non-negligible, but these extreme cases show how observing a larger number of ions can probe the relative importance of each effect. Note especially the three ions with $\zeta \approx 0.25$ (He II, O V, and Si VIII). The relative difference in their widths is a decisive probe of the strength of Coulomb collisions, since purely collisionless cyclotron resonance would have predicted identical perpendicular heating rates. Note also Si XI, which has $\zeta = 0.36$ (extremely close to the charge-to-mass ratio of $Mg^{9+}$). A large difference between the Si XI line width and the Mg X line width implies strong (and very frequency-localized) wave damping at the $O^{6+}$ resonance, as discussed above. As the number of observable ions increases, the constraints that can be placed on the physical processes acting in the corona grow stronger.

ACKNOWLEDGMENTS

This work is supported by the National Aeronautics and Space Administration under grant NAG5-11420 to the Smithsonian Astrophysical Observatory, by Agenzia Spaziale Italiana, and by the Swiss contribution to the ESA PRODEX program.

REFERENCES

Aschwanden, M. J., Acton, L. W., 2001, ApJ, 550, 475

Axford, W. I., McKenzie, J. F., 1992, in Solar Wind Seven, ed. E. Marsch & R. Schwenn (New York: Pergamon), 1

Banaszkiewicz, M., Axford, W. I., McKenzie, J. F., 1998, A&A, 337, 940

Barnes, A., 1992, Rev. Geophys., 30, 43

Barnes, A., Gazis, P. R., Phillips, J. L., 1995, GRL, 22, 3309

Bieber, J. W., Wanner, W., Matthaeus, W. H., 1996, JGR, 101, 2511

Collier, M. R., Hamilton, D. C., Gloeckler, G., Bochsler, P., Sheldon, R. B., 1996, GRL, 23, 1191

Cranmer, S. R., 1998, ApJ, 508, 925

Cranmer, S. R., 2000, ApJ, 532, 1197

Cranmer, S. R., 2001a, in Radio Frequency Power in Plasmas, 14th Topical Conference, ed. T. K. Mau & J. deGrassie (AIP Conf. Proc. 595), 25

Cranmer, S. R., 2001b, JGR, 106, 24937

Cranmer, S. R., 2002a, in Multi-wavelength Observations of Coronal Structure and Dynamics, ed. P. C. H. Martens & D. Cauffman (Elsevier, COSPAR Colloq. Ser.), in press

Cranmer, S. R., 2002b, Space Sci. Rev., in press

Cranmer, S. R., Field, G. B., Kohl, J. L., 1999a, ApJ, 518, 937

Cranmer, S. R., Kohl, J. L., Noci, G., et al., 1999b, ApJ, 511, 481

David, C., Gabriel, A. H., Bely-Dubau, F., Fludra, A., Lemaire, P., Wilhelm, K., 1998, A&A, 336, L90

Doschek, G. A., Feldman, U., Laming, J. M., Schühle, U., Wilhelm, K., 2001, ApJ, 546, 559

Dowdy, J. F., Jr., Rabin, D., Moore, R. L., 1986, Solar Phys., 105, 35

Esser, R., Edgar, R. J., 2000, ApJ, 532, L71

Esser, R., Fineschi, S., Dobrzycka, D., Habbal, S. R., Edgar, R. J., Raymond, J. C., Kohl, J. L., Guhathakurta, M., 1999, ApJ, 510, L63

Esser, R., Habbal, S. R., 1997, in Cosmic Winds and the Heliosphere, ed. J. R. Jokipii, C. P. Sonett, & M. S. Giampapa (Tucson: U. Arizona Press), 297

Feldman, W. C., Marsch, E., 1997, in Cosmic Winds and the Heliosphere, ed. J. R. Jokipii, C. P. Sonett, & M. S. Giampapa (Tucson: U. Arizona Press), 617

Fineschi, S., Gardner, L. D., Kohl, J. L., Romoli, M., Noci, G., 1998, Proc. S.P.I.E., 3443, 67

Giordano, S., Antonucci, E., Noci, G., Romoli, M., Kohl, J. L., 2000, ApJ, 531, L79

Goldreich, P., Sridhar, S., 1997, ApJ, 485, 680

Habbal, S. R., Leer, E., 1982, ApJ, 253, 318

Hassler, D. M., Dammasch, I. E., Lemaire, P., Brekke, P., Curdt, W., Mason, H. E., Vial, J.-C., Wilhelm, K., 1999, Science, 283, 810

Hassler, D. M., Wilhelm, K., Lemaire, P., and Schühle, U., 1997, Solar Phys., 175, 375

Hefti, S., Grünwaldt, H., Ipavich, F. M., et al., 1998, JGR, 103, 29697

Hollweg, J. V., 1986, JGR, 91, 4111

Hollweg, J. V., 1999, JGR, 104, 24781

Hollweg, J. V., 2000, JGR, 105, 7573

Hollweg, J. V., Isenberg, P. A., 2002, JGR, 107, in press

Hu, Y.-Q., Habbal, S. R., Li, X., 1999, JGr, 104, 24819

Hundhausen, A. J., 1972, Coronal Expansion and Solar Wind (Berlin: Springer-Verlag)

Isenberg, P. A., Hollweg, J. V., 1983, JGR, 88, 3923

Ko, Y.-K., Fisk, L. A., Geiss, J., Gloeckler, G., Guhathakurta, M., 1997, Solar Phys., 171, 345

Kohl, J. L., Esser, R., Cranmer, S. R., et al., 1999, ApJ, 510, L59

Kohl, J. L., Noci, G., Antonucci, E., et al., 1997, Solar Phys., 175, 613

Kohl, J. L., Noci, G., Antonucci, E., et al., 1998, ApJ, 501, L127

Kohl, J. L., Reeves, E. M., Kirkham, B., 1978, in New Instrumentation for Space Astronomy, ed. K. van der Hucht & G. Vaiana (New York: Pergamon), 91

Krieger, A. S., Timothy, A. F., Roelof, E. C., 1973, Solar Phys., 29, 505

Leamon, R. J., Matthaeus, W. H., Smith, C. W., Wong, H. K., 1998, ApJ, 507, L181

Leer, E., Holzer, T. E., Flå, T., 1982, Space Sci. Rev., 33, 161

Leer, E., Hansteen, V. H., Holzer, T. E., 1998, in Cyclical Variability in Stellar Winds, ed. L. Kaper & A. W. Fullerton (Berlin: Springer-Verlag), 263

Lemaire, J., Pierrard, V., 2001, Astrophys. Space Sci., 277, 169

Li, X., 1999, JGR, 104, 19773

Li, X., Habbal, S. R., Kohl, J. L., Noci, G., 1998, ApJ, 501, L133

Lie-Svendsen, Ø., Hansteen, V. H., Leer, E., Holzer, T. E., 2002, ApJ, 566, 562

Markovskii, S. A., 2001, ApJ, 557, 337

Marsch, E., 1999, Space Sci. Rev., 87, 1

Marsch, E., Mühlhäuser, K.-H., Schwenn, R., Rosenbauer, H., Pilipp, W., Neubauer, F. M., 1982, JGR, 87, 52

Matthaeus, W. H., Goldstein, M. L., Roberts, D. A., 1990, JGR, 95, 20673

Matthaeus, W. H., Zank, G. P., Oughton, S., Mullan, D. J., Dmitruk, P., 1999, ApJ, 523, L93

McKenzie, J. F., Banaszkiewicz, M., Axford, W. I., 1995, A&A, 303, L45

Miralles, M. P., Cranmer, S. R., Panasyuk, A. V., Romoli, M., Kohl, J. L., 2001a, ApJ, 549, L257

Miralles, M. P., Cranmer, S. R., Kohl, J. L., 2001b, ApJ, 560, L193

Narain, U., Ulmschneider, P., 1990, Space Sci. Rev., 54, 377

Narain, U., Ulmschneider, P., 1996, Space Sci. Rev., 75, 453

Nolte, J. T., Krieger, A. S., Timothy, A. F., et al., 1976, Solar Phys., 46, 303

Ofman, L., Nakariakov, V. M., DeForest, C. E., 1999, ApJ, 514, 441

Parker, E. N., 1963, Interplanetary Dynamical Processes, (New York: Interscience Publishers)

Parker, E. N., 1991, ApJ, 372, 719

Parker, E. N., 2001, JGR, 106, 15797

Peter, H., Judge, P. G., 1999, ApJ, 522, 1148

Phillips, J. L., Feldman, W. C., Gosling, J. T., Scime, E. E., 1995, Adv. Space Res., 16 (9), 95

Pneuman, G. W., 1986, Space Sci. Rev., 43, 105

Priest, E. R., Foley, C. R., Heyvaerts, J., Arber, T. D., Mackay, D., et al., 2000, ApJ, 539, 1002

Ruzmaikin, A., Berger, M. A. 1998, A&A, 337, L9

Seely, J. F., Feldman, U., Schühle, U., Wilhelm, K., Curdt, W., Lemaire, P., 1997, ApJ, 484, L87

Shebalin, J. V., Matthaeus, W. H., Montgomery, D., 1983, J. Plasma Phys., 29, 525

Spitzer, L., Jr., 1962, Physics of Fully Ionized Gases, 2nd ed. (New York: Wiley)

Tu, C.-Y., Marsch, E., 1995, Space Sci. Rev., 73, 1

Tu, C.-Y., Marsch, E., 1997, Solar Phys., 171, 363

Tu, C.-Y., Marsch, E., 2001, JGR, 106, 8233

Tu, C.-Y., Marsch, E., Wilhelm, K., Curdt, W., 1998, ApJ, 503, 475

Velli, M., 2001, Astrophys. Space Sci., 277, 157

Waldmeier, M., 1957, Die Sonnenkorona, vol. 2 (Basel: Verlag Birkhäuser)

Waldmeier, M., 1975, Solar Phys., 40, 351

Wang, Y.-M., Sheeley, N. R., Jr., 1990, ApJ, 355, 726

Wilcox, J. M., 1968, Space Sci. Rev., 8, 258

Withbroe, G. L., Kohl, J. L., Weiser, H., Munro, R. H., 1982, Space Sci. Rev., 33, 17

Zhou, Y., Matthaeus, W. H., 1990, JGR, 95, 14881

Zirker, J. B., ed., 1977 Coronal Holes and High-speed Wind Streams (Boulder: Colorado Assoc. Univ. Press)